\documentclass[pss,fleqn]{w-art}
\usepackage{times}
\usepackage{w-thm}

\usepackage{graphicx}

\providecommand{\unit}[1]{\,\mbox{#1}}

\DeclareMathAlphabet{\mathvec}{OT1}{cmr}{bx}{sl}
\SetMathAlphabet{\mathvec}{bold}{OT1}{cmr}{bx}{sl}

\newcommand{\ua}{\uparrow}
\newcommand{\da}{\downarrow}

\newcommand{\nund}{{n_{\ua},n_{\da}}}

\begin{document}
\pagespan{1}{}



\title[Ringlike structures in the $n_{2D}$--$B$ $\rho_{xx}$ diagram of quantum Hall systems]{Ringlike
structures in the density--magnetic-field $\rho_{xx}$ diagram of
two-subband quantum Hall systems}


\author[Ferreira]{Gerson J. Ferreira\footnote{Corresponding author: e-mail: {\sf gersonjr@ifsc.usp.br}}\inst{1}
\address[\inst{1}]{Instituto de F\'{\i}sica de S\~ao Carlos, Universidade de S\~{a}o Paulo,\\
Caixa Postal 369, 13560-590 S\~ao Carlos, SP, Brazil}}
\author[Freire]{Henrique J. P. Freire\inst{1}}
\author[Egues]{J. Carlos Egues\inst{1,2}
\address[\inst{2}]{Department of Physics and Astronomy, University of Basel, CH-4056 Basel, Switzerland}
}
\begin{abstract}
Motivated by recent experiments [Zhang \textit{et al.}, Phys. Rev.
Lett. \textbf{95}, 216801 (2005) and Ellenberger \textit{et al.},
cond-mat/0602271] reporting novel ringlike structures in the
density--magnetic-field ($n_{2D}$--$B$) diagrams of the
longitudinal resistivity $\rho_{xx}$ of quantum wells with two
subbands, we investigate theoretically here the magneto-transport
properties of these quantum-Hall systems. We determine $\rho_{xx}$
via both the Hartree and the Kohn-Sham self-consistent schemes plus
the Kubo formula. While the Hartree calculation yields
diamond-shaped structures in the $n_{2D}$--$B$ diagram, the
calculation including exchange and correlation effects (Kohn-Sham)
more closely reproduces the ringlike structures in the experiments.
\end{abstract}

\maketitle                   






\section{Introduction}

Quantum Hall systems, i.e., two-dimensional electron gases (2DEGs)
under strong magnetic fields, exhibit a variety of novel physical
phenomena \cite{sarma-pinc}. Besides the widely known quantum Hall
effects, these systems also display quantum Hall ferromagnetism --
an spontaneously formed spin-polarized many-body state arising from
the interplay of the Zeeman, Coulomb, and thermal energy scales
within the highly-degenerate Landau states of the 2DEG \cite{quinn}.

Quantum Hall ferromagnetism has been experimentally investigated via
magneto-transport measurements of the longitudinal $\rho_{xx}$ and
Hall $\rho_{xy}$ resistances in both the integer and fractional
quantum Hall regimes. A crucial ingredient in these experiments is
the occurrence of opposite-spin Landau level crossings
\cite{fang-stiles} near the Fermi energy. At these crossings the
system may find it energetically favorable to spin polarize thus
reducing its energy due to the Pauli-Coulomb exchange interaction.
Resistance spikes (or ``anomalous'' peaks) are found in $\rho_{xx}$
at these crossings. Corresponding dips or bumps in $\rho_{xy}$ are
sometimes reported. Most importantly, these features in $\rho_{xx}$
and $\rho_{xy}$ are hysteretic in many cases thus signaling
Ising-like ferromagnetism
\cite{Piazza99,Jungwirth01,Jaroszynski02,Freire04condmat}.

Recent experiments in two-subband quantum Hall systems
\cite{Zhang05,Ellenberger06} have found interesting ringlike
structures (Fig. 1(a), reproduced from Ref. \cite{Zhang05}) in the
density--magnetic-field ($n_{2D}$--$B$) diagram of $\rho_{xx}$.
The authors of Ref. \cite{Zhang05} claim that these unusual
structures arise from quantum Hall ferromagnet states in the system,
while those of Ref. \cite{Ellenberger06} explain their ringlike
structures in terms of a phenomenological single-particle picture.
In both experiments crossings of Landau levels from distinct
confined subbands occur at the Fermi level.

\begin{figure}[htb]
\begin{center}
    \includegraphics[width=\textwidth, keepaspectratio=true]{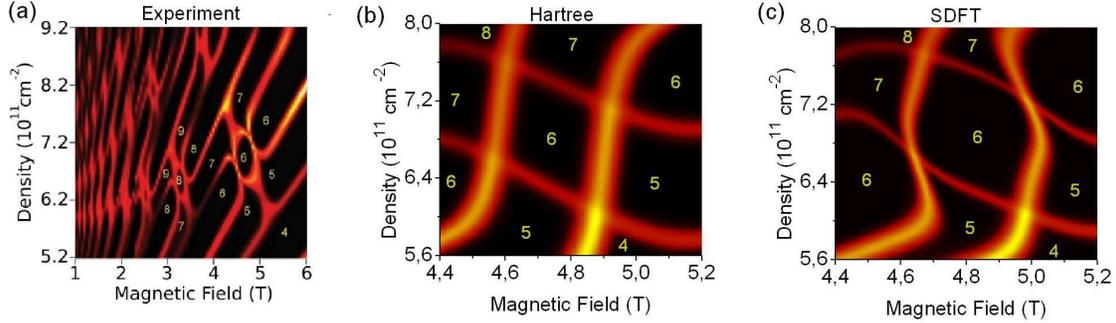}
    \caption{Experimental (a) (adapted from Ref. \cite{Zhang05}) and calculated density--magnetic-field $\rho_{xx}$
    diagrams within the Hartree approximation (b) and the SDFT/LSDA (c). Only the
    ringlike structure corresponding to filling factor $\nu=6$ is shown in (b) and (c).}
    \label{fig:1}
\end{center}
\end{figure}

Here we theoretically investigate the magneto-transport properties
of quantum-Hall systems with two occupied subbands. We consider a
modulation-doped GaAs/AlGaAs quantum well, similar to that of Ref.
\cite{Zhang05},\cite{both}. We determine the subband structure of
the well via both (i) the Hartree approach and (ii) the Spin Density
Functional Theory (SDFT) \cite{Hohenberg64} [within the Local Spin
density Approximation (LSDA)] implemented via the usual Kohn-Sham
scheme \cite{Kohn65}. We obtain the magneto-conductivity tensor (and
its inverse, the magneto-resistivity) from the Kubo formula. We find
diamond-shaped structures [Fig. 1(b)] in the $n_{2D}$--$B$
diagram within the Hartree approximation, while the SDFT-LSDA
calculation yields smoother structures [Fig. 1(c)] more closely
resembling the ringlike shapes in Refs. \cite{Zhang05,Ellenberger06}
[e.g., Fig. 1(a)]. We do not find any hysteresis or discontinuities
in the DFT-LSDA Landau-level fan diagram, which would indicate
easy-axis (Ising) quantum Hall ferromagnetism in the system
\cite{Piazza99,Jungwirth01,Jaroszynski02,Freire04condmat}. However,
we can not rule out the possibility for easy-plane ferromagnetism.

\section{Self-consistent approach}

\textit{Hartree and SDFT/LSDA calculations.} Within the usual
implementation of the SDFT/LSDA formulation in the context of the
effective mass approximation, we self-consistently solve the
Kohn-Sham equations for the conduction electrons in an external
potential. Because of the translational symmetry in the $xy$ plane
of the 2DEG, the problem is separable \cite{Freire04condmat}. For the $z$ direction we have
two (spin up and down) coupled one-dimensional Schr\"odinger
equations
\begin{equation}
\left[ -\frac{\hbar^2 }{2m}\frac{d }{d z^{2}}+v_{eff}^{\sigma
_{z}}\left( z;[n_{\uparrow },n_{\downarrow }]\right) \right] \chi
_{i}^{\sigma _{z}}\left( z\right) =\varepsilon _{i}^{\sigma
_{z}}\chi _{i}^{\sigma _{z}}\left( z\right) \, ,
\label{eqSch}
\end{equation}
with $m$ being the effective mass, $i=1,2\ldots $ the subband index,
and $v_{eff}^{\sigma_{z}}(z)$ the effective single-particle
potential. While the heterostructure potential confines the
longitudinal motion, the perpendicular magnetic field collapses the
transversal motion into Landau levels with energies
$\varepsilon_n=(n+1/2)\hbar\omega_c$, $\omega_c=eB/m$, each of which
with a macroscopic degeneracy $eB/h$ ($\sim 10^{11}\unit{cm}^{-2}$).
Therefore, the Kohn-Sham single-particle energy of each electronic
level is
\begin{equation}
  \varepsilon _{i,n}^{\sigma _{z}} =
       \varepsilon _{i}^{\sigma _{z}}
                + \left( n + \frac{1}{2}\right) \hbar \omega_c
        + \frac{\sigma _{z}}{2}g\mu_{B}B \, .
\label{eqLL}
\end{equation}
where $g\mu_B B \sigma_z/2$ is the ordinary Zeeman term and $g$ the
effective Land\'e factor. The spin-dependent effective potential in
Eq.~(\ref{eqSch}) consists of two main contributions: the confining
potential of the heterostructure $v_{c}(z)$ and the Coulomb
potential which in the Kohn-Sham/SDFT scheme is split into Hartree
$v_{H}(z,[n])$ and an exchange-correlation (XC) $v_{xc}(z,[\nund])$
contributions. The one-dimensional effective potential in the
$z$-direction then reads
\begin{equation}
  v^{\sigma_z}_{eff}(z,[\nund]) = v_c(z)
                                + v_H(z,[n])
                    + v^{\sigma_z}_{xc}(z,[\nund]) \, .
\label{eqVeff}
\end{equation}
In our simulations, the structural potential $v_c(z)$ represents the
square quantum well investigated in Ref.~\cite{Zhang05}. The Hartree potential
is calculated by solving the Poisson's equation taking into account
the electronic charge density \emph{and} the positive charge profile
of the depleted modulation-doped regions. For the XC potential we
use the Vosko-Wilk-Nusair (VWN) \cite{Vosko80} parameterization of the LSDA.

Since $v_H$ and $v_{xc}$ are both functionals of $n_\ua$ and
$n_\da$, which are calculated from the Kohn-Sham orbitals
$\chi_i^{\sigma_{z}}(z)$, Eqs.~(\ref{eqSch}) and (\ref{eqVeff}) are
solved self-consistently. The densities, and any other observable of
interest, are calculated using a Gaussian broadened density of
states (DOS) weighed by the Fermi function (\textit{aufbau}
principle). Each level is characterized by a width
$\Gamma^{\sigma_z}$ with an electron density $eB/h$. Note that the
Hartree calculation is obtained from the above self-consistent
scheme by simply turning off the XC contribution.

\textit{Magneto-transport.} We determine the magneto-resistances
from the magneto-conductivity tensor
($\mathbf{\rho}=\mathbf{\sigma}^{-1}$), calculated within the
self-consistent Born-approximation model of Ando and Uemura
\cite{Ando74,Ando74III}. For short-range scatterers,
\begin{equation}
  \sigma_{xx} = \frac{4e^2}{\hbar}
     \int\limits_{-\infty}^{\infty}
     \left( -\frac{\partial f(\varepsilon)}{\partial \varepsilon}  \right)
     \sum_{i,n,\sigma_z} \left(n+\frac{1}{2}\right)
     \exp \left[ - \left( \frac{\varepsilon-\varepsilon_{i,n}^{\sigma_z}}
       {\Gamma_{\mathrm{ext}}} \right)^2 \right] \mathrm{d} \varepsilon
\end{equation}
where $\sigma_{xy} = e n_{2D}/B + \Delta\sigma_{xy}$ with $\Delta
\sigma_{xy}$ being a small correction \cite{Ando74,Ando74III}. We
model the extended Landau states by a Gaussian DOS
$g_{\mathrm{ext}}(\varepsilon)$ of width
$\Gamma_\mathrm{ext}=0.25\unit{meV}$ \cite{Prange81}. Next we
present our results.

\section{Results and discussions}

\emph{Parameters.} We consider a GaAs/Al$_{0.3}$Ga$_{0.7}$As
modulation-doped quantum well \cite{Zhang05} with a width $L=240\unit{\AA}$
and two symmetric $\delta$-doped layers with planar density
$n_{2D} = 10^{12}\unit{cm}^{-2}$ (which is identical to the electron
density in our simulations), set back from the well by $240\unit{\AA}$. We
use an effective g-factor $g = -0.44$, $m = 0.063$, a potential well
offset $\sim 236.43\unit{meV}$ \cite{Madelung84,Vurgaftman01} and $T =
340\unit{mK}$ \cite{Zhang05} in our simulations.

\textit{Hartree calculation.} First of all we note that in a
Hartree-type calculation the Zeeman interaction is the only source
of the spin splitting of the energy levels. Moreover, this splitting
increases linearly with $B$ and does not depend on the electron
density $n_{2D}$. Using standard parameters for GaAs wells,
\textit{e.g.} $g = -0.44$ and $m = 0.063$, our Hartree calculation
yields $n_{2D}$--$B$ $\rho_{xx}$ diagrams showing small
diamond-shaped structures (this shape follows from the linear
dependence of the spin splitting with $B$) while the experiment
\cite{Zhang05} exhibits much larger ring-like structures [Fig.
\ref{fig:1}(a)]. In order to generate structures with an area
comparable to the experimental rings, we need to use an enhanced $g
\sim -2.1$ in our Hartree simulation. This value is consistent with
the experimental estimate of Ellenberger \textit{et al.}
\cite{Ellenberger06} who correctly suggests (as our SDFT/LSDA
calculation below shows) that this enhancement is due to
exchange-correlation effects. Figure \ref{fig:1}(b) shows the
$\nu=6$ diamond. The small area of the diamonds calculated with the
standard parameters suggests that the Zeeman effect is not the only
source of spin splitting in the experiments. This is indeed the
case, as we find in the SDFT/LSDA calculation.

\textit{SDFT/LSDA calculation.} Here, in contrast to the Hartree
case, the spin-dependent exchange-correlation potential
$v^{\sigma_z}_{xc}(z,[\nund])$ in the Kohn-Sham equations gives rise
to a density-dependent spin splitting, not linear in $B$. Figure
\ref{fig:1}(c) shows the $\nu=6$ structure in the $n_{2D}$--$B$
diagram, calculated via the SDFT/LSDA with standard parameters for
GaAs wells. We can see a smoother structure which more closely
resembles the ring-like structures in the experiment of Ref.
\cite{Zhang05} [Fig. \ref{fig:1}(a)]. We have also calculated the
full range $n_{2D}$--$B$ $\rho_{xx}$ diagram and find that it
is very similar to the experimental one, Fig. \ref{fig:1}(a) \cite{Neck}.
Therefore, the SDFT-type calculation with the standard GaAs
parameters results in more rounded structures with areas and shapes
similar to those reported in Refs. \cite{Zhang05} and
\cite{Ellenberger06}.

\textit{Quantum Hall ferromagnetism.} The above results seem to
corroborate the conjecture in Ref. \cite{Zhang05} that the ringlike
shapes of the structures should be due to density-dependent
exchange-correlation effects. However, our SDFT/LSDA simulations do
not show ferromagnetic phase transitions. The Landau fan diagram and
the resistivities ($\rho_{xx}$ and $\rho_{xy}$) do not show any
hysteresis or discontinuities which would signal an easy-axis
(Ising-like) ferromagnetic transition
\cite{Piazza99,Jungwirth01,Jaroszynski02,Freire04condmat}. In
addition, the spin-polarization at the center of the rings are too
small to be understood as an easy-axis ferromagnetic phase. However,
we cannot rule out easy-plane (\textit{xy}) ferromagnetism. In fact,
an early experimental investigation \cite{Muraki01} has reported on
the observation of both easy-axis and easy-plane quantum-Hall
ferromagnetism in similar systems.

\section{Summary}

We have calculated the $n_{2D}$--$B$ diagram of $\rho_{xx}$
within both the Hartree and the SDFT/LSDA approaches (plus the Kubo
formula) in two-subband quantum Hall systems. While our Hartree
calculation with an artificially enhanced $g$ factor of $-2.1$
yields diamond-shaped structures, the calculation with density- and
spin-dependent exchange-correlation effects and the standard
$g=-0.44$ value, more closely reproduces the ringlike shapes seen
experimentally.

\begin{acknowledgement}
GJF thanks X. Zhang for providing some experimental details. JCE
acknowledges useful discussions with K. Ensslin. This work was
supported by CNPq and FAPESP.
\end{acknowledgement}


\end{document}